\documentclass[journal]{IEEEtran}
\usepackage{cite}
\usepackage{amsmath,amssymb,amsfonts}
\usepackage{algorithmic}
\usepackage{graphicx,color}
\usepackage{textcomp}
\usepackage{gensymb}
\def\BibTeX{{\rm B\kern-.05em{\sc i\kern-.025em b}\kern-.08em
		T\kern-.1667em\lower.7ex\hbox{E}\kern-.125emX}}

\usepackage[caption=false,font=normalsize,labelfont=sf,textfont=sf]{subfig}

\begin{document}

	
	\title{Additive Frequency Diverse Active Incoherent Millimeter-Wave Imaging}
	
	\author{Jorge R. Colon-Berrios,~\IEEEmembership{Graduate Student Member,~IEEE,} and Jeffrey A. Nanzer,~\IEEEmembership{Senior Member,~IEEE}

	\thanks{The authors are with the Department of Electrical and Computer Engineering, Michigan State University, East Lansing, MI 48824 USA (email: \{colonbe1, nanzer\}@msu.edu).}
	} %
	\maketitle
	
	\begin{abstract}
		
		We present an approach for improving spatial frequency sampling in active incoherent millimeter-wave (AIM) imaging systems using frequency diversity. AIM imaging relies on active transmission of spatio-temporally incoherent signals to illuminate a scene, from which interferometric Fourier-domain imaging can be implemented using a sparse receiving antenna array. One of the benefits of Fourier domain imaging is the sparsity of the receiving array, which can form images with equivalent resolution to traditional filled beamsteering arrays, but with a small fraction of the elements. The hardware reduction afforded by the sparse array often leads to an undersampled Fourier space, where even though image formation is possible, the image reconstruction may be degraded when viewing complex objects. To address this challenge without requiring additional receiver channels, we explore the use of frequency diversity in the illuminating and receiving systems. Fourier domain spatial frequency samples are determined by the electrical spacing and rotation of the receiving elements, thus by changing the frequency the sampled spatial frequencies also change. We implement an additive technique where the spatial frequency samples are summed prior to Fourier transform image formation. Importantly, because the system is active, a consistent signal-to-noise ratio is maintained across all frequencies, which may not be possible in traditional passive Fourier-domain imagers. We implement the approach in a 24-element receiver array oriented in a circular pattern combined with a four-element noise transmitter. Image improvement is evaluated through simulation on various test scenes utilizing a structural similarity index measure (SSIM) to determine the quality of image reconstruction. Experiments were conducted in 1~GHz frequency steps ranging from 37--40~GHz on scenes consisting of reflecting spheres and cylinders, demonstrating a reduction of spurious signals in the resultant additive image.
	\end{abstract}
	
	\thispagestyle{plain} 
	\pagestyle{plain} 

	%
	\IEEEpeerreviewmaketitle
	\section{Introduction}
	
	Millimeter-wave frequencies are widely used for sensing applications because the wavelengths of the radiation in the 30-300~GHz range are short enough to provide reasonable image resolution while also being long enough to propagate through various obscurants like garment materials and smoke with minimal attenuation~\cite{currie1987principles,nanzer2012microwave}. Millimeter-wave radiation also does not penetrate human skin, making it a safer technique than systems like X-ray imagers~\cite{6758} and has been increasingly applied to applications such as contraband detection, medical imaging, and remote sensing~\cite{yujiri2003passive,4337827,942570}. Millimeter-wave imaging systems can be classified as either passive or active. Passive millimeter-wave imaging systems detect the intrinsic thermal radiation emitted by people and can be used to detect hidden objects by the differences in their emissivity~\cite{1492659}. The drawback of such systems is the significant signal gain necessary in the receiver since thermal radiation in the millimeter-wave band is exceedingly small. However, since thermal radiation is spatially and temporally incoherent, passive imagers can use Fourier-domain interferometric imaging which can generate images with sparse antenna apertures, reducing the hardware burden compared to typical filled apertures like phased arrays or focal plane arrays~\cite{Thompson2001,6688267,6048955,898661,6305002,7685, 8416691}.
	\begin{figure*}[t!]
		\centering
		\includegraphics[width=0.75\textwidth]{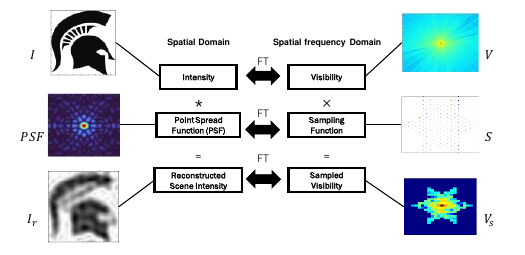}
		\caption{Overview of the Fourier domain imaging process. AIM imaging operates by measuring information in the spatial frequency domain, which is the Fourier transform of the scene intensity $I$. The spatial frequency information, called the visibility $V$, is multiplied by the sampling function $S$, which is determined by the layout of the receiver array. The inverse Fourier transform of the sampling function is the point spread function (PSF). The product of the visibility and the sampling function is the sampled visibility $V_s$, which is inverse Fourier transformed to obtain the reconstructed scene intensity $I_r$. The reconstructed intensity can also be viewed as the convolution of the intensity and the PSF.}
	\label{processing}
\end{figure*}
Active imagers overcome the low signal power limitation by illuminating the scene and capturing the reflected signals, which leads to increased signal-to-noise ratio (SNR) and reduce requirements on the receiver hardware~\cite{Hunt310,10.1117/12.488198}. However, active systems typically operate coherently by transmitting known signals which reflect off the scene with high levels of spatial and temporal coherence. Traditional radar-based imaging leverages this coherence using matched filter processing, for example, which increases the SNR. However, to achieve good angular resolution, a filled aperture is often needed, such as a phased array, focal plane array, or a large mechanical reflector, each of which has drawbacks in terms of size, weight, power, and cost compared to Fourier-domain interferometric imagers which use sparse apertures with far fewer hardware requirements. Computational imaging approaches have been developed to reduce the hardware burden through various processing approaches, however the processing complexity often increases the image formation time, making fast imaging challenging and furthermore requiring significant computational resources~\cite{8972939,8055576,8036228}. We recently developed an active incoherent millimeter-wave (AIM) imaging approach that leverages the transmission of noise waveforms to illuminate the scene, providing higher SNR at the receivers while maintaining the spatial and temporal incoherence necessary for Fourier domain image formation~\cite{8458190,9079644}. AIM imaging enables faster Fourier domain image reconstruction than passive imagers~\cite{8654605}, and more control over the imaging process, which can improve image formation. Fourier domain interferometric imagers collect spatial frequency samples via the cross-correlation of signals from pairs of receivers in the array; the spatial sampling function is thus limited by the number of elements, and few elements may result in poor image quality.

In this work, we explore an approach to improving spatial frequency sampling in AIM imaging based on frequency diversity. The spatial frequency samples collected by the receiving array are determined by the electrical separation of pairs of antennas; thus, by changing the frequency of the measurement, new spatial frequency samples can be obtained. Importantly, since AIM imaging uses active transmission, the SNR of the received signals can be controlled more than in passive imagers, where the received signal power is dependent on the frequency-dependent emissivity of objects~\cite{nanzer2012microwave}, leading to more consistent sampling across frequency bands. Previously we explored the use of frequency diversity by multiplicatively combining reconstructed images across different frequencies~\cite{10221679}. While this approach mitigates the impact of frequency-dependent interference, it has the tendency to enhance strong targets and diminish weak targets. In this paper, we take a more comprehensive approach where we use frequency diversity to change the sampling function of the receiver, increasing the number of spatial frequency samples collected by the imager. The samples are combined in the spatial Fourier domain, and then processed via inverse Fourier transform to reconstruct the image. We explore the approach through simulation and measurement with a $k_a$-band imager with a 24-element receiver in a circular array format. Fourier domain samples are captured at center frequencies of 37-40 GHz in 1 GHz increments. We characterize the image improvement through simulation for a set of test scenes, and conduct measurements of scenes consisting of reflecting spheres and cylinders, demonstrating the improvement in image formation of the additive AIM technique compared to images formed at single frequencies.


\begin{figure*}[t!]
	\centering
	\includegraphics[width=1\textwidth]{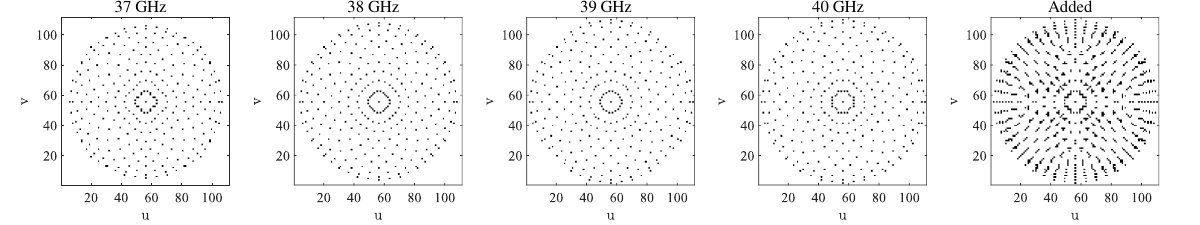}
	\caption{Sampling functions of the 24-element at frequencies of 37, 38, 39, 40~GHz and the additive sampling function, which yields significantly more spatial frequency samples than an individual frequency sampling function.}
	\label{SF}
\end{figure*}

\section{\textcolor{black}{Active Fourier Domain Imaging}}
\label{Section.Theory}

Fourier domain imaging relies on measurements of the mutual coherence of electromagnetic signals captured by antennas in the receiving array~\cite{Thompson2001}. By cross-correlating the received signals, a sample of the information in the spatial Fourier domain is captured; this information is called the scene \textit{visibility} $V$, and if the signals emanating from the scene are spatially and temporally incoherent, the scene intensity $I$ is related to the visibility through a Fourier transform by
\begin{equation}\label{eq.vis}
	V(u,v) = \iint\limits_{-\infty}^{+\infty}I(\alpha,\beta)e^{j2\pi(u\alpha+v\beta)}d\alpha d\beta
\end{equation}
where $u$ and $v$ are spatial frequencies and $\alpha = \sin\theta \cos\phi$ and $\beta=\sin\theta \sin\phi$ are direction cosines.

Reconstructing the scene intensity $I$ is accomplished through an inverse Fourier transform of the scene visibility. However, the ability to sample the visibility is limited by the receiver hardware. In order to capture visibility samples the signals received in the array are cross-correlated pairwise, forming a correlation interferometer with each antenna pair, thereby generating a sampling function determined by the set of antenna pairs in the array which can be given by
\begin{equation}\label{eq.sf}
	S(u,v) = \sum_{n}^{N}\sum_{m}^{M}\delta(u-u_n)\delta(v-v_m)\mathrm{,}
\end{equation}
where $\delta(\cdot)$ is the delta Dirac function and $N\times M$ is the total number of samples, and $u=\frac{D_x}{\lambda}$ and $v=\frac{D_y}{\lambda}$ where $D_x$ and $D_y$ are the separations of the antenna pairs in the $x$ and $y$ dimensions, and $\lambda$ is the wavelength of the received radiation. Since the visibility sample is determined by the relative separation and angle between pairs of antennas, dense sampling functions are generally obtained using sparse arrays with diversity in the separations and relative angles. 

The reconstructed intensity is represented by
\begin{equation}\label{eq.recon}
	I_r(\alpha,\beta) = \iint\limits_{-\infty}^{+\infty}V(u,v)S(u,v)e^{-j2\pi(u\alpha+v\beta)}dudv
\end{equation}
where the product of the visibility and the sampling function defines a sampled visibility that is a finite set of points captured by the array. Since the array captures visibility in discrete points, the reconstructed intensity can be given by
\begin{equation}\label{eq.ir}
	I_r(\alpha,\beta) = \sum_{n}^{N}\sum_{m}^{M}V(u_n, v_m)e^{-j2\pi\left(u_n\alpha+v_m\beta\right)}
\end{equation}
\begin{figure*}[t!]
	\centering
	\includegraphics[width=1\textwidth]{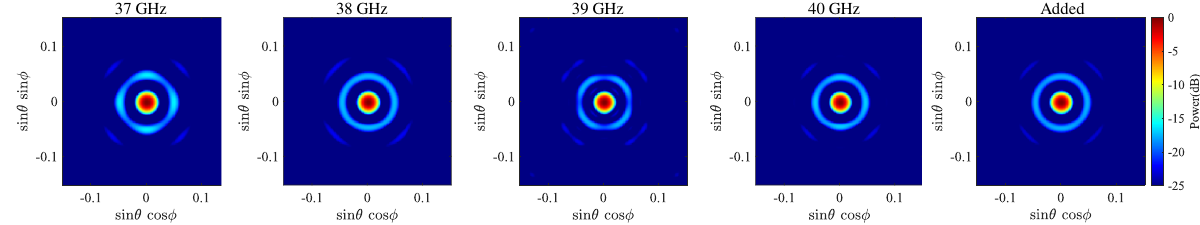}
	\caption{Theoretical PSF for each frequency subband as well as for the additive sampling function. The additive case yields a more uniform PSF with lower sidelobes.}
	\label{TheoPSF}
\end{figure*}

The spatio-temporal incoherence condition is met in passive Fourier-domain imaging systems since such systems capture intrinsic thermally-generated electromagnetic radiation emitted by objects in the scene, which is naturally noise-like and incoherent across space and time. However, the power contained in these natural emissions are exceedingly low for terrestrial objects in the microwave and millimeter-wave bands~\cite{nanzer2012microwave}, thus receivers with significant gain (often on the order of 120 dB) with long integration times are required to achieve sufficient SNR for imaging. These requirements lead to receivers that are complex and costly and also form images at relatively slow rates due to the need for long integration time. To overcome these limitations, active transmission can be used, however the signals must still satisfy the spatio-temporal incoherence condition. AIM imaging accomplishes this with a set of noise transmitters which, if spaced outside the span of the receiving array and transmitting independent noise, generate sufficiently noise-like signals reflecting off the scene to allow for Fourier domain reconstruction.
Image reconstruction can then proceed as detailed above, however the receiver gain and integration time can be significantly lower than in passive systems; image formation at a rate of hundreds of frames per second are possible~\cite{9450727}.


Fig.~\ref{processing} gives an overview of the imaging process. While the image formation process proceeds in the spatial Fourier domain, it is helpful to understand the spatial domain representation of the process. The scene intensity $I$ represents the information to be recovered; its Fourier transform is $V$. The receiving array defines the sampling function $S$ in the spatial frequency domain, and the product of the sampling function and visibility is the sampled visibility $V_s$. The spatial domain equivalent is shown on the left side of the figure. The Fourier transform of the sampling function is the point spread function (PSF); an ideal sampling function covering all spatial frequencies would result in a PSF that is a delta function. The PSF from a typical sampling function has a finite width main lobe and sidelobes, which result in a degradation in the reconstructed image intensity $I_r$, which is the convolution of the image intensity and the PSF.

\section{Additive Frequency Diverse AIM Imaging}


The sampling function is limited by the physical setup of the receiving array since the number of unique samples obtained by the imager depends on the number of unique antenna pairs. The spatial frequency sample is determined entirely by the electrical separation, or baseline, of the antennas in the pair and the relative angle (i.e., the electrical separation in the $x$ and $y$ dimensions as noted in the previous section). Arrays with significant uniformity, such as layouts on a rectangular grid like traditional phased arrays, thus yield a considerable number of redundant spatial frequency samples since there are a large number of repeated baselines. Such redundancy can be helpful for calibration purposes~\cite{9079644}, but does not make the most efficient use of the hardware in the array. Designing arrays that more efficiently sample the Fourier domain leads to more spatial frequency samples and better image reconstruction.

Alternatively, dynamics can be used to increase the number of spatial frequency samples. Since the spatial frequencies $u$ and $v$ are dependent on the electrical separation of the antenna pairs, physical motion of the elements is sometimes a viable option for increasing the number of spatial frequency samples. Physically moving the elements in the array in subsequent measurements was common in early radio astronomy application, while the use of Earth's rotation to change the projected baselines and thus increase spatial frequency samples over time has become standard~\cite{Thompson2001}. While these dynamics take place on the scale of hours or days, faster physical motion has also been explored for measurements on the order of seconds with as few as two receiving elements~\cite{9395397,9660363}. 


Physical motion of the antenna elements in the receiving array is not always feasible, particularly for fast imaging applications, but since the spatial frequencies depend on the electrical separation, the wavelength can also be dynamically changed to increase the number of spatial frequency samples. Here we explore the use of multi-frequency measurements within practical limits of millimeter-wave hardware to improve image reconstruction quality. An important benefit of frequency diversity in AIM imaging is the ability to more directly control the received signal SNR than in passive imaging systems. Passive imagers detect thermally-generated electromagnetic radiation, which is dependent on the emissivity of materials in the scene, which is generally a function of frequency and may change appreciably across the band of interest in multi-frequency systems~\cite{nanzer2012microwave}. By actively transmitting noise signals in AIM imaging, the reflected signal power can be more accurately controlled, leading to a more consistent SNR across the entire band of interest, leading to more reliable and robust image reconstruction.

Previously we explored a multiplicative multifrequency imaging approach where images were formed at separate subbands and subsequently multiplied~\cite{10221679}. The resultant images effectively remove frequency-dependent spurious signals such as narrowband interferers, however the multiplicative process has the limitation that strong targets tend to be enhanced in intensity while weak targets tend to be reduced in intensity. In this work, we explore a more robust and general additive approach where combination of the subband samples is accomplished in the visibilty domain. In particular, wavelength-dependent sampling functions $S_{\lambda_k}$ are generated at $K$ carrier frequencies, after which the additive sampled visibility is given by
\begin{equation}\label{Sampled Visibility}
	V_a = \sum_{k}^{K} V S_{\lambda_k}
\end{equation} 
The reconstructed image intensity is then given by
\begin{equation}\label{eq.iradd}
	I_r(\alpha,\beta) = \sum_{n}^{N}\sum_{m}^{M}V_a(u_n, v_m)e^{-j2\pi\left(u_n\alpha+v_m\beta\right)}
\end{equation}

We evaluate the approach through simulation in this section and through experimentation in the following section. We consider a $k_a$-band AIM imaging system with a 24-element receiver operating at subbands centered at 37--40~GHz in 1~GHz increments ($K=4$) with a noise bandwidth of $\Delta f = $~50~MHz at each subband. We consider a circular O-shaped receiving array with a radius of 101~mm and elements positioned $15\degree$ of spacing. This simulated setup is consistent with the experimental setup in the following section.

Fig.~\ref{SF} shows the sampling function $S_{\lambda_k}$ for each of the subbands. The sampling functions capture a total of 289 (at 37~GHz, 38~GHz, and 40~GHz) or 281 (at 39~GHz) unique samples , the difference being due to discretization of the $u$--$v$ space based on $\lambda/2$ sampling. Fig.~\ref{SF} shows the additive sampling function $S_a = \sum{S_{\lambda_k}}$ which yields a total of 935 unique spatial frequency samples, an increase by a factor of 3.24. Again, due to the discretized spatial frequency grid, the total number of points in $S_a$ is not precisely the sum of the number of points in each individual sampling function, as some samples fall in the same spatial frequency bin. The addition of more subbands will result in more samples, however the differences between the frequency subbands must be large enough to ensure that subsequent sampling function yields points at different bins in the discretized grid.

The PSFs for each individual sampling function $S_{\lambda_k}$ as well as the PSF of the additive sampling function $S_a$ are shown in Fig~\ref{TheoPSF}. The main lobe of the individual PSFs remains largely consistent due to the similarities of the sampling functions, however the sidelobe structure of the individual PSFs are change across frequency. These small changes impact image reconstruction, as will be seen later in the experimental section. The additive PSF shows a more angularly consistent principal sidelobe.

\section{Analysis of Image Quality Through Simulated Scene Reconstruction}

We evaluated the impact on image quality through simulation for a set of test scenes. Several scenes were designed, some with low spatial frequency content, and others with a range of low to high spatial frequency content. We assume each image to be contained within the unambiguous field of view of the array. Reconstructions were simulated at individual frequencies in the range of 35-45~GHz in 1~GHz increments, along with the reconstruction of the added visibility from all 11 subbands. In order to compare the performance or quality of image reconstruction the metric similar structure index metric (SSIM) was utilized. SSIM is an image quality metric that considers the structural information, contrast, and luminescence between the recovered image and the reference image and has a range of [-1, 1] where 1 represents identical recover and reference images, 0 and -1 indicate no similarity and perfect anti-correlation, respectively~\cite{1284395}. Because of the similarities of the sampling functions as shown in the previous section, it was not anticipated that large differences would be observed between individual frequencies and the resultant additive image; however, as shown later, the experimental results showed large differences. 

\begin{figure}[t!]
	\centering
	\includegraphics[width=0.5\textwidth]{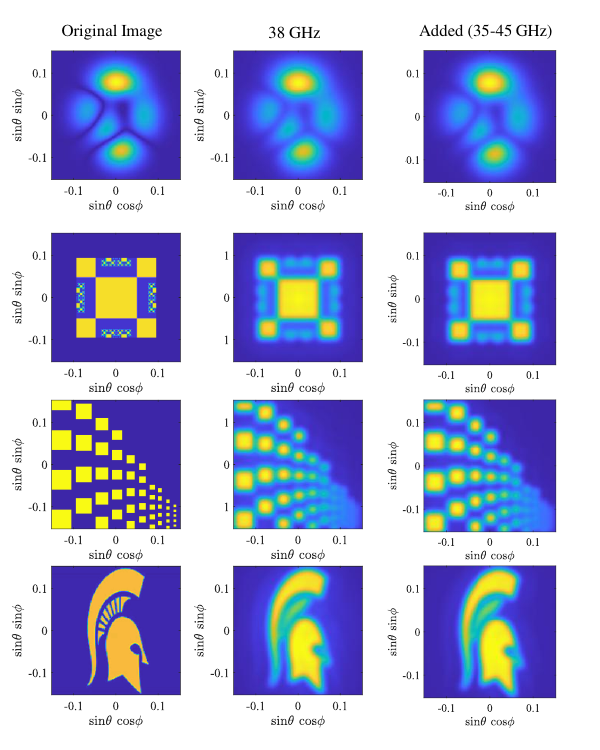}
	\caption{Sample images used for the simulation analysis. The left column is the original scene intensity; the middle column is the reconstructed image at one frequency subband (38 GHz); the right column is the reconstruction with the additive case. Four test images were used: S1 is a low spatial frequency image; S2 contains squares with exponentially decreasing size, leading to more high-frequency content due the edges; S3 consists of squares of linearly decreasing size, generating a broad spatial spectral image; and S4 is the Michigan State University Spartan helmet. The SSIM values and the improvement in SSIM for the additive case are given in Table~\ref{SSIM}.}
	\label{analysis}
\end{figure}

Fig.~\ref{analysis} shows the test images, an example of a single frequency reconstruction at 38~GHz, and the added reconstruction using the visibility obtained from adding the information from all subbands. The first test scene (S1) case, where the spatial change in intensity is smooth, resulting in most of the information residing in the low spatial frequency region. The second scene (S2) is a square fractal image with numerous edges and small shapes, resulting in significant more content in the high spatial frequency region, however due to the fractal nature of the scene, the shape sizes decrease exponentially, leading to the majority of the information residing at lower spatial frequencies. The third scene (S3) is a series of squares of decreasing size which has a more continuous distribution of energy across low and high spatial frequencies. Finally, the fourth scene (S4) is a Michigan State University Spartan helmet, which also has a diversity of spatial frequency content, and represents a complex real-world type of image.
\begin{table*}[t!]\caption{SSIM Values Calculated From Fig.~\ref{analysis}}\label{SSIM}
	\begin{center}
		\begin{tabular}{c|c|c|c|c|c|c|c|c|c|c|c|c|c}
			\hline\hline
			& 35 GHz & 36 GHz & 37 GHz & 38 GHz & 39 GHz & 40 GHz & 41 GHz & 42 GHz & 43 GHz & 44 GHz & 45 GHz & \textbf{Added} & Increase* \\ \hline
			S1 & 0.714 & 0.695 & 0.711 & 0.711 & 0.734 & 0.713 & 0.730 & 0.748 & 0.748 & 0.710 & 0.698 & \textbf{0.863} & 16.65\% \\ 
			S2 & 0.182 & 0.183 & 0.197 & 0.199 & 0.212 & 0.207 & 0.210 & 0.216 & 0.220 & 0.202 & 0.193 & \textbf{0.346 }& 41.65\% \\ 
			S3 & 0.194 & 0.168 & 0.183 & 0.183 & 0.198 & 0.191 & 0.198 & 0.204 & 0.207 & 0.190 & 0.178 & \textbf{0.278} & 31.53\% \\ 
			S4 & 0.367 & 0.339 & 0.364 & 0.362 & 0.379 & 0.369 & 0.374 & 0.388 & 0.391 & 0.361 & 0.343 & \textbf{0.522} & 29.69\% \\ \hline\hline
		\end{tabular}
	\end{center}
	*Percentage increase of Added SSIM over average SSIM across all subbands.
\end{table*}
The SSIM for each case was calculated by using the normalized original scene as a reference and the normalized reconstruction as the recovered image. Table~\ref{SSIM} shows the SSIM for each image at each individual frequency subband as well as the additive image. The additive approach increases the SSIM in each case. Note that because of the similarities of the sampling functions, appreciable improvements in image quality manifest in sidelobe structure, leading to appreciable improvements in SSIM, although the basic shape appears similar to the human eye. 

The ability to reconstruct the scenes accurately with high SSIM is dependent on the resolution of the imaging system. S1 contains mostly low-spatial frequency information, which can be reconstructed with a low resolution imaging system, thus the average SSIM is higher for S1 than the other cases. However, the important aspect of this work is how significantly the SSIM can be improved through the additive imaging technique compared to a single subband reconstruction. Thus, the percentage improvement in SSIM between the average SSIM across all subbands and the SSIM of the reconstructed additive image was calculated.
The scenes with the most energy concentrated at low spatial frequencies (S1 and S2) show improvements of approximately 17\% and 41\% respectfully, the broad-frequency image S3 showed the greatest improvement of more than 31\%, and the Spartan helmet more than 29\%. Because the sampling functions are similar across the individual and added images, the resultant resolution of the reconstructed images is also similar, thus the major information gained through the additive technique manifests at middle and high spatial frequencies, where reduction of the sidelobe structure in the PSF has a larger impact. 

\begin{figure}[t!]
\centering
\includegraphics[width=0.4\textwidth]{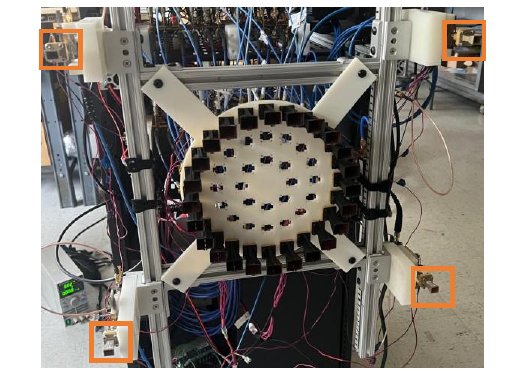}
\caption{Optical Image of the system, consists of four transmitters (orange squares) and 24 receivers in a circular array layout.}
\label{System}
\end{figure}
\section{Experimental Evaluation}

\subsection{System Setup}

\begin{figure}[t!]
\centering
\includegraphics[width=0.4\textwidth]{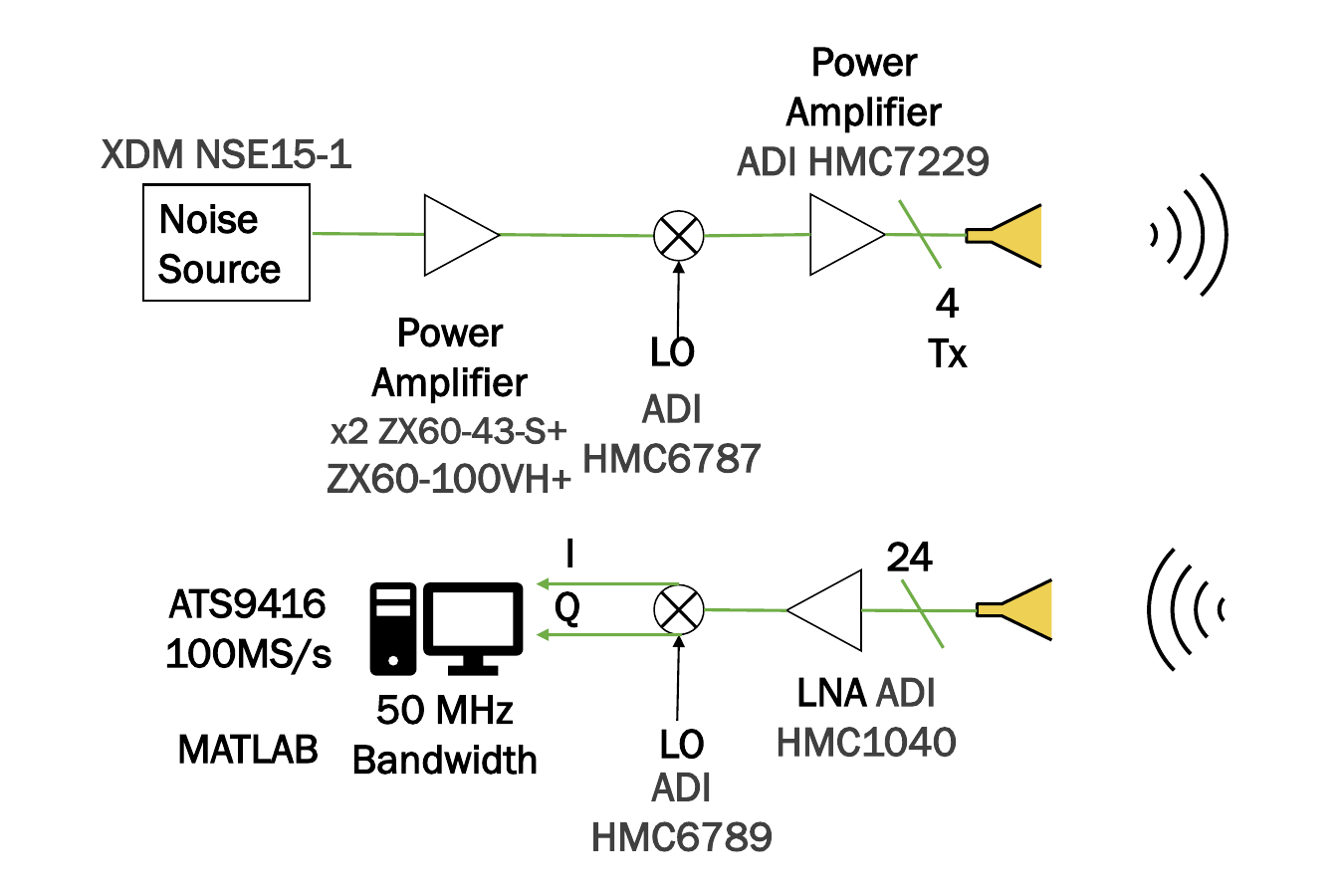}
\caption{Schematic of the AIM imaging system.}
\label{Schematic}
\end{figure}

In order to demonstrate the performance of the approach experiments were performed using the AIM system shown in Fig.~\ref{System}, the schematic is shown in Fig.~\ref{Schematic}. The imager consists of four independent noise transmitters and twenty four receivers. The transmitters were placed at a wider baseline than the receivers to ensure that the signals incident on the scene were decorrelated at a spatial resolution finer than that of the receiver. Each transmitter used a separate calibrated baseband noise source, the signals from which were first amplified using three amplifiers: two ZX60-43-S+ and one ZX60-100VH+ with gains of 23.1~dB and 26~dB respectfully in cascade and then upconverted to the millimeter-wave carrier frequencies using Analog Devices (ADI) HMC6787 upconverters. Before transmission, the millimeter-wave noise signals are amplified again using ADI HMC7229 power amplifiers with 24~dB of gain. The subband carrier frequency was generated by changing the upconverter local oscillator (LO) using a Keysight N5183A signal generator. The upconverters included an integrated frequency doubler, and the subsequent carrier frequencies were hopped between 37~GHz, 38~GHz, 39~GHz and 40~GHz using LO frequencies of 18.5~GHz, 19~GHz, 19.5~GHz, and 20~GHz, respectively. 
Importantly, the chosen subbands were all within the operational bandwidths of the system hardware. This means that all subbands can be measured by only changing the LO frequency and does not require any other hardware changes.

\begin{figure}[t!]
\centering
\includegraphics[width=0.5\textwidth]{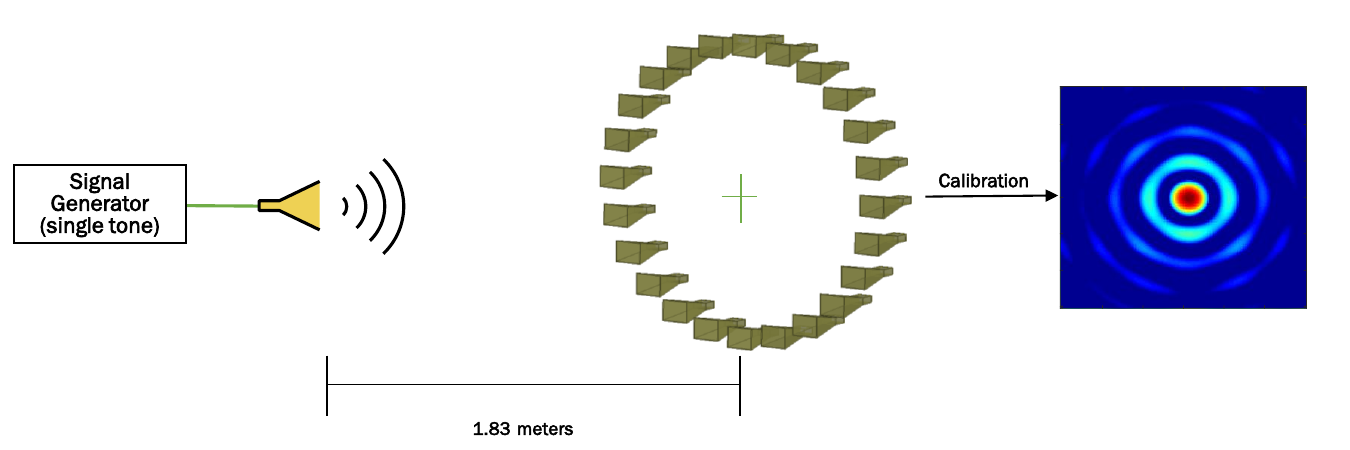}
\caption{Calibration was performed using a transmitting antenna acting as a strong point sources. Weights are calculated via optimization for each of the subbands.}
\label{calibration}
\end{figure}

\begin{figure*}[t!]
\centering
\includegraphics[width=\textwidth]{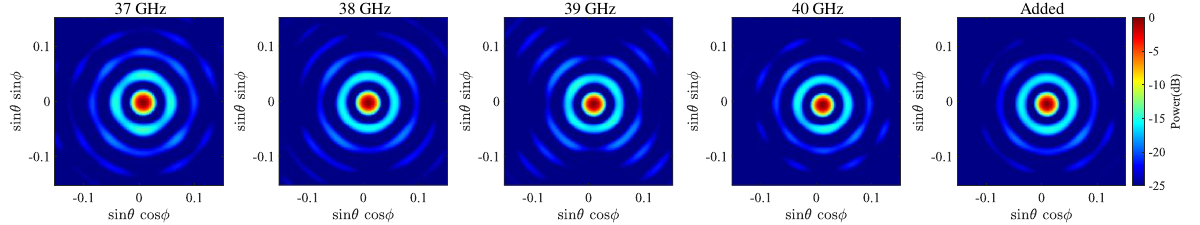}
\caption{Measured PSF for each frequency subband as well as for the additive sampling function. The additive case yields a more uniform PSF with lower sidelobes. The sidelobes are stronger in the measured case compared to the simulation case likely because of hardware nonidealities.}
\label{measPSF}
\end{figure*}
The upconverted noise signals were transmitted to the scene and scattered back towards the receiving array. Since the transmitter noise sources were independent, the signals incident on the scene were spatially and temporally incoherent, allowing Fourier image reconstruction as described in Section~\ref{Section.Theory}. 
The signals captured at the receiving array were initially amplified using ADI HMC1040 low-noise amplifiers with 20~dB of gain at these frequencies, and then quadrature downconverted using ADI HMC6789 downconverters, after which the 48 I/Q baseband signals were digitized by three ATS9416 samplers that were hosted in a computer. The sample rate on each channel was 100MS/s. All signal processing and image reconstruction was implemented in a host computer using MATLAB.

\subsection{System Calibration}

Calibration of the receiving array was performed at each subband frequency using an active continuous-wave transmitter at the subband center frequency, as show, in Fig.~\ref{calibration}. The calibration source was positioned at 1.83 meters from the receiver. In this work a similar approach to~\cite{9079644} was applied for calibrating the system, however instead of using a point target we use an active transmitter to ensure high SNR. 
Since the target is known to be a point target at the center of the array, following~\cite{90000} an optimization is performed on the amplitude and phase weights of each receiver based on the known form of the point target. This yielded 24 weights at each subband, and a total of 96 weights for all four subbands.

The measured PSFs for each subband and the additive PSF are shown in Figs.~\ref{measPSF}, which match closely to the theoretical plots. We note that nonidealities in the hardware lead to increased sidelobes in the experiments, as secondary, tertiary, and quaternary sidelobes are more prominent in the measured PSFs. 
Since at each frequency the components of the system have different gains, the resultant visibility samples have different amplitudes across the subbands.
This can cause certain measurements to dominate the reconstruction if not calibrated. To address this, we normalize the visibiltiy samples per subband prior to superposition, such that each band contributes equally to the resultant additive image.

\begin{figure}[t!]
\centering
\includegraphics[width=0.5\textwidth]{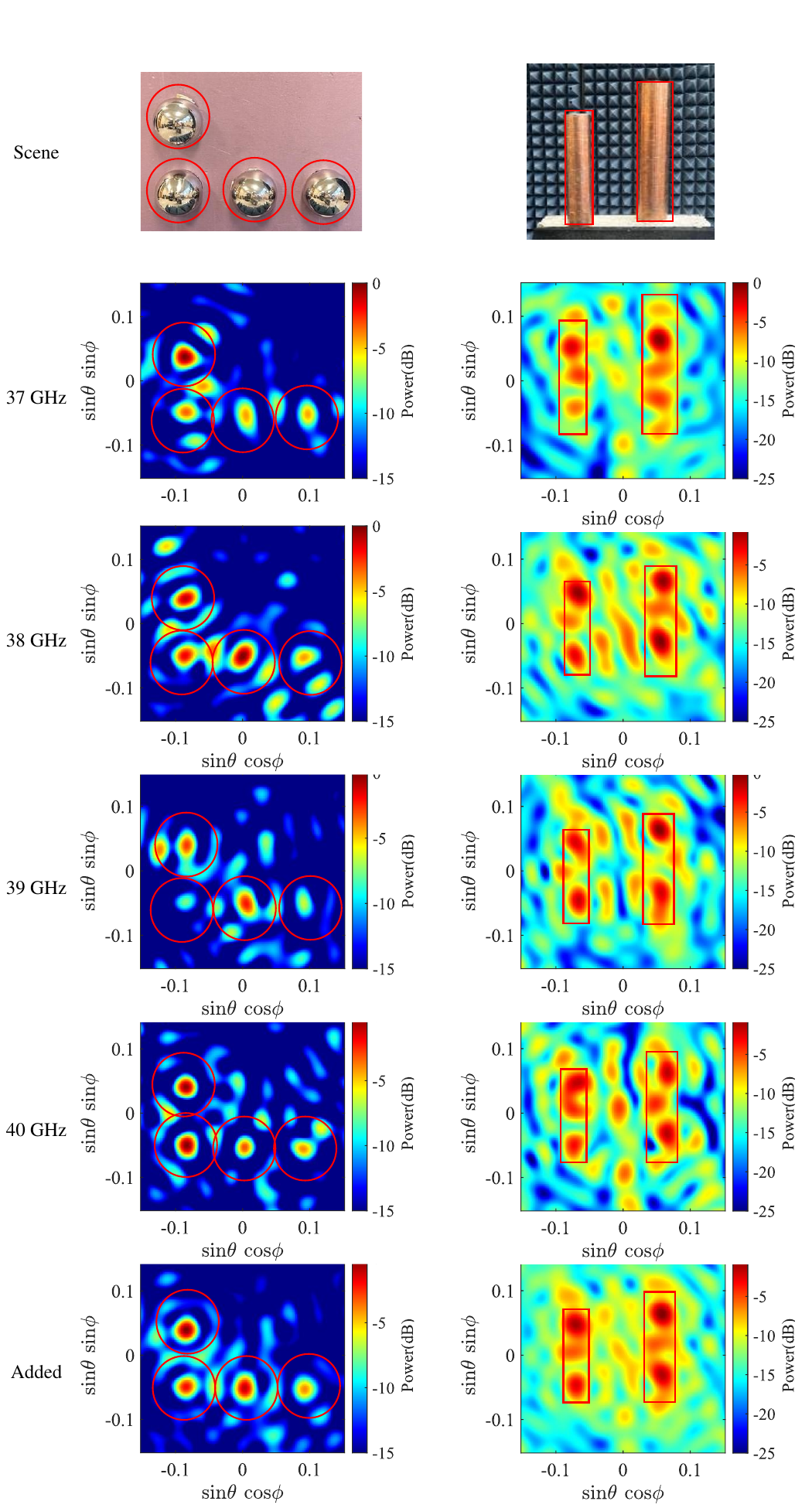}
\caption{Experimental imaging results for two targets. The left columns shows the reconstructed images of four metallic spheres at each subband and the additive case; red circles indicate the locations of the spheres. It can be seen that in most of the individual subband cases some of the targets are missed, while numerous spurious artifacts are also present. The additive case recovers each of the targets and significantly reduces the artifacts, leading to a more accurate reconstruction. The right column shows the reconstructed images from two conducting cylinders; the red rectangles indicate the locations of the cylinders. Again the individual subbands produce images with significant spurious artifacts, while the additive case yields a more accurate reconstruction with significantly fewer spurious artifacts.}
\label{Results}
\end{figure}

\subsection{Imaging Experiments}

Two different scenes were created for experimental evaluation. The first consisted of 
four metallic spheres in an L formation. Each sphere was 10~cm in diameter and had a separation of 15~cm center to center. The target was placed at 1.5~m from the receive array of the system; the system and the targets remained stationary throughout the measurement. On the left side of Fig.~\ref{Results} we can see the target as well as the image reconstruction of the four spheres at each frequency along with the additive image. It is apparent that nonidealities in the experimental system led to significant variation in image reconstruction quality across subbands. In some cases, the response from a sphere is absent; in other cases strong artifacts manifest at angles where no sphere is located. 
The additive case shows an image with the four spheres in the correct locations and with minimal spurious signals and sidelobes throughout the rest of the image, leading to a more accurate reconstruction of the scene.

The second scene consisted of two cylinders covered in copper tape placed at a distance of 1.8~m from the receiving array. The cylinder on the right was 47~cm tall with a diameter of 10~cm, while the one on the left was 37~cm of height with a diameter of 8~cm. The spacing between the targets was 31~cm. 
The cylindrical shape poses a greater image reconstruction challenge than the spheres due to specularity, which can cause issues when reconstructing images with arrays~\cite{10745128}. This is apparent in the individual subband images in Fig.~\ref{Results}, where the location of the cylinders is apparent in most images, but the shape is difficult to discern. The multifrequency additive approach has the effect of reducing the impact of specularity issues since they are frequency dependent. The resulting additive image shows a much cleaner, more accurate reconstruction of the cylinders, supporting this concept. 


\section{Conclusion}

We explore the use of frequency diversity and additive spatial frequency sampling to improve image quality in AIM imaging systems. In contrast to passive Fourier domain imaging systems, AIM imaging has the benefit of controlled transmission across all bands, leading to better SNR at all frequencies and a more reliable superposition of the measurements in each subband. The improvement in image quality was demonstrated through simulation using a set of complex scenes mimicking the performance of a practical 24-element interferometric imaging array. Measurements of two scenes using a $k_a$-band AIM imager with four noise transmitters and a 24-element receiver in an O-array layout demonstrated the effectiveness of the additive technique in practical imaging applications of complex reflecting scenes. Because the approach relies only on changing the LO frequency, without requiring any other hardware changes to measure different subbands, the technique can be applied without significant modifications or cost, leading to improved image reconstruction quality in Fourier domain millimeter-wave imaging applications.

\ifCLASSOPTIONcaptionsoff
\newpage
\fi

\bibliographystyle{IEEEtran}
\bibliography{IEEEabrv,reference_bib}

\end{document}